\documentclass[aps,prl,
 ]{revtex4}
\usepackage{graphicx}
\usepackage{color}
\usepackage{slashed}
\usepackage{amsmath}
\newcommand\bef{\begin{figure}}
\newcommand\eef[1]{\label{fg:#1}\end{figure}}
\newcommand\beq{\begin{equation}}
\newcommand\eeq[1]{\label{#1}\end{equation}}
\newcommand\beqa{\begin{eqnarray}}
\newcommand\eeqa[1]{\label{#1}\end{eqnarray}}
\newcommand\bet{\begin{table}}
\newcommand\eet[1]{\label{tb:#1}\end{table}}

\newcommand\fgn[1]{Figure \ref{fg:#1}}
\newcommand\eqn[1]{eq.\ (\ref{#1})}

\newcommand\tr{{\rm Tr\/}}

\newcommand\ie{{\sl i.e.\/}}

\newcommand{\N}{{\cal N}}

\newcommand{\bilin}[1]{\overline\psi{#1}\psi}

\newcommand\ppbar{\langle\overline\psi\psi\rangle}
\newcommand\F{{\scriptscriptstyle F}}

\begin{document}

\title{Real time warm pions from the lattice using an effective theory}
\author{Sourendu\ \surname{Gupta}}
\email{sgupta@theory.tifr.res.in}
\affiliation{Department of Theoretical Physics, Tata Institute of Fundamental
         Research,\\ Homi Bhabha Road, Mumbai 400005, India.}
\author{Rishi\ \surname{Sharma}}
\email{rishi@theory.tifr.res.in}
\affiliation{Department of Theoretical Physics, Tata Institute of Fundamental
         Research,\\ Homi Bhabha Road, Mumbai 400005, India.}
\begin{abstract}
Lattice measurements provide adequate information to fix the parameters
of long distance effective field theories in Euclidean time. Using such a
theory, we examine the analytic continuation of long distance correlation
functions of composite operators at finite temperature from Euclidean to
Minkowski space time. There are two definitions of mass in each regime;
in Euclidean these are the screening and pole masses. The analytic
continuation of these mass parameters to real time is non-trivial.
This is in contrast to the situation at zero temperature.\\
TIFR/TH/19-16
\end{abstract}
\maketitle

The computation of the thermodynamics of quantum field theories is under
good control using the Euclidean formulation and non-perturbative lattice
computations. However the analytic continuation to real time Minkowski
quantities remains an open problem, despite decades of attempts to chip
away at it. The first attempt to extract a transport coefficient from
lattice computations was made more than three decades ago \cite{wyld}.
However, it wasn't until fifteen years later that it was realized
that more control was needed on the non-perturbative structure of the
spectral density function \cite{gert}.  Despite advances in weak-coupling
expansions \cite{amy}, the introduction of new methods \cite{stats},
and many lattice computations \cite{many}, the extraction of real-time
dynamics at finite temperature from lattice computations is far from
becoming a routine measurement. This is a matter of concern, because
there have been improved measurements of many flow variables in heavy-ion
collision experiments \cite{flow}, and it seems possible to start on
the extraction of transport coefficients from data.

Using an effective field theory (EFT) to model finite temperature physics
in QCD, we have been able to describe accurately the long-distance
behaviour of static correlation functions of flavoured axial currents
in lattice QCD \cite{Gupta:2017gbs}. In the limit of vanishing quark
mass, these currents are conserved. As a result, in real-time dynamics
there should be diffusive transport of flavoured axial charge. Even
for physically relevant light quark masses, since the pion mass is
small compared to the typical QCD scale, one might expect this charge
to be a slow, albeit non-conserved, mode in a QCD fluid. As a result,
it is interesting to examine what the EFT approach tells us about the
connection between Euclidean and real-time dynamics. Here we report a
first step to this: the non-trivial connection between pion correlators
in Euclidean and real time dynamics.

We constructed a sequence of two effective field theory models at finite
temperature.  The existence of a special frame where the heat bath is at
rest implies a lack of boost invariance in the Lagrangian, as a result
of which the global symmetries are of spatial rotations, apart from a
SU(2)$\times$SU(2) chiral-flavour symmetry.  The starting model is of
self-interacting quarks, the Minkowski version of which has the Lagrangian
\beqa
\nonumber
  L &=& -d^3T_0\bilin{} + \bilin{\slashed\partial^0} 
      + d^4 \bilin{\slashed\nabla} 
      + \frac{d^{61}}{T_0^2} \left[(\bilin{})^2 
            + (\bilin{i\gamma_5\tau^a})^2 \right]
      + \frac{d^{62}}{T_0^2} \left[(\bilin{\tau^a})^2 
            + (\bilin{i\gamma_5})^2 \right] 
\\ \nonumber &&
      + \frac{d^{63}}{T_0^2} (\bilin{\gamma_0})(\bilin{\gamma^0}) 
      + \frac{d^{64}}{T_0^2} (\bilin{i\gamma_i})(\bilin{i\gamma^i})
      + \frac{d^{65}}{T_0^2} (\bilin{\gamma_5\gamma_0})
                             (\bilin{\gamma_5\gamma^0}) 
      + \frac{d^{66}}{T_0^2} (\bilin{i\gamma_5\gamma_i})
                             (\bilin{i\gamma_5\gamma^i})  
\\ \nonumber &&
      + \frac{d^{67}}{T_0^2} \left[
      (\bilin{\gamma_0\tau^a})
      (\bilin{\gamma^0\tau^a}) 
      + (\bilin{\gamma_5\gamma_0\tau^a})
        (\bilin{\gamma_5\gamma^0\tau^a})\right]
      + \frac{d^{68}}{T_0^2} \left[
              (\bilin{i\gamma_i\tau^a})(\bilin{i\gamma^i\tau^a}) 
            + (\bilin{i\gamma_5\gamma_i\tau^a})(\bilin{i\gamma_5\gamma^i\tau^a})\right]  
\\ &&
+ \frac{d^{69}}{T_0^2} \left[(\bilin{iS_{i0}})(\bilin{iS^{i0}}) 
            + (\bilin{S_{ij}\tau^a})^2 \right]
      + \frac{d^{60}}{T_0^2} \left[(\bilin{iS_{i0}\tau^a})(\bilin{iS^{i0}\tau^a}) 
            + (\bilin{S_{ij}})^2 \right].
\eeqa{qlag}
where $\slashed\partial^0=\gamma_0\partial^0$ and $\slashed\nabla =
\gamma_j\partial^j$ where $j$ runs over all spatial indices and $\tau^a$
are the generators of flavour SU(2). We use the metric conventions of
\cite{wein}. This theory is defined with a cutoff, $T_0$, which we will
choose so that physics at the temperatures of interest can be described
by the theory.  For later simplicity in writing formul{\ae}, we introduce
the notation $\N=4N_cN_f$ for the number of components of quark fields
and $m_0=d^3T_0$.

Since this is similar to the Nambu-Jona-Lasinio (NJL) model \cite{njl},
known techniques \cite{klevansky} can be used to first analyze the mean
field theory. In the chiral limit, $m_0=0$, chiral symmetry is broken
spontaneously, a non-vanishing quark condensate, $\ppbar$, is produced,
and a second-order chiral symmetry restoring phase transition (at a
temperature that we chooose to be $T_0$) is found. The single combination
of the dimension-6 couplings,
\beq
 \lambda=(\N+2)d^{61}-2d^{62}-d^{63}+3d^{64}+d^{65}-3d^{66}-3d^{69}+3d^{60},
\eeq{lambda}
appears in the subsequent physics that we discuss \footnote{This formula
corrects a typographical error in \cite{Gupta:2017gbs}}. In the Euclidean
theory these conclusions followed from the computation of the free
energy. In real time they come from a self-consistent solution of the
one-loop Schwinger-Dyson equation for the quark propagator using real time
perturbation theory (we follow the conventions of \cite{Kobes:1984vb}). In
both cases we use dimensional regularization with $\overline{MS}$
subtraction. Since the expressions for the quark condensate are exactly
the same in the two computations, the phase structure of the theory can
be computed in either the Euclidean or real-time formalism \cite{Landsman:1986uw}.

At finite $m_0$, the quark mass explicitly breaks chiral
symmetry. Nevertheless, a remnant of spontaneous chiral symmetry breaking
appears as a large value of the condensate, giving an effective quark mass
$m=m_0+\Sigma$, where $\Sigma=2\lambda\ppbar/T_0^2$. As the temperature
is increased, $\Sigma$ crosses over to a small value. This is exactly
what was found in the Euclidean computation of \cite{Gupta:2017gbs}.

The fixing of the parameters in the action of \eqn{qlag} was done by
matching one-loop expressions for the axial current correlators to the
lattice measurements of \cite{brandt}. The process was simple; since
the axial symmetry is broken, small fluctuations around the mean field
have the quantum numbers of the pion. After integrating over the quark
fields, one could obtain an effective pion action at longer distance
scales. In the chiral limit the axial vector current is conserved; the
conservation law is broken only by the parameter $d^3$ in the action.
Using the one-loop version of the PCAC relation at finite temperature,
the long-distance static axial current correlator could be parametrized
in terms of the coupling constants $d^3$ and $d^4$. The convention
that $T_0$ is the chiral transition temperature $T_c$ fixes $\lambda$,
and the value of $T_0$ was then inferred from matching the cross over
temperature, $T_{co}$, at finite $m_0$ simultaneously with the matching
of axial current correlators \cite{Gupta:2017gbs}.

The specific question that we ask here is how the parameters which
govern the long-distance part of the pion correlation function can be
analytically continued to real time.  At $T=0$ the pion mass and decay
constant can be measured from the long-distance properties of correlation
functions measured in Euclidean lattice computations. These Euclidean
measurements simply give the value of the corresponding real time
quantities. However, at finite temperature this is a fraught question.
Since the pion is a composite operator, analytic continuation of its
long-distance properties requires knowledge of a thermal spectral density
function. If this is as straightforward as at $T=0$, then the Euclidean
pion effective theory that one obtains can be simply taken over to real
time, and computation of the axial flavour charge diffusion constant
should be straightforward. On the other hand, if correlations of the
composite pion operator are to be treated with the same care as the
correlation functions of, say, the energy-momentum tensor, then getting
the effective pion theory in real-time may be non-trivial.  Since we
have a tractable theory in which the pion is composite, namely that in
\eqn{qlag}, we can try to answer this question by analytic continuation
of that theory. This is what we demonstrate next.

Since the axial symmetry is broken, small amplitude collective
fluctuations around the mean field can be parametrized using a
Hubbard-Stratanovich transformation,
\beq
 \psi\to\exp\left[\frac{i{\bf\pi}\gamma^5}{2f}\right]\,\psi
\eeq{hst}
with a three component field ${\bf\pi}=\pi^a\tau^a$,
and a constant $f$ with the dimension of mass. Introducing
this parametrization into the fermion action and expanding to
second order in $\pi$
gives a coupled model of quarks and mesons
\beq
 L_c = -\bilin{\left[d^3T_0\left(1+i\gamma^5\frac\pi f
           -\frac{\pi^2}{2f^2}\right)-\slashed\partial
           -\frac i{2f}\gamma^5\slashed\partial\pi\right]} + \cdots
\eeq{coupled}
where $\slashed\partial=\partial^0+d^4\slashed\nabla$, and the terms of
dimension-6 have not been written out.  The pion appears as an auxiliary
field, and hence has no kinetic term in $L_c$.  The path integral
over pion fields can be damped by adding a quadratic term of the form
$\epsilon\pi^a\pi^a$ to $L_c$.  We will examine the causal correlator
for pions taking them to be external fields only, so the $11$-component
of the real-time propagator, ${\cal D}_{11}$ \cite{Kobes:1984vb} is all
that is necessary.

\bef
\begin{center}
\includegraphics[scale=0.40]{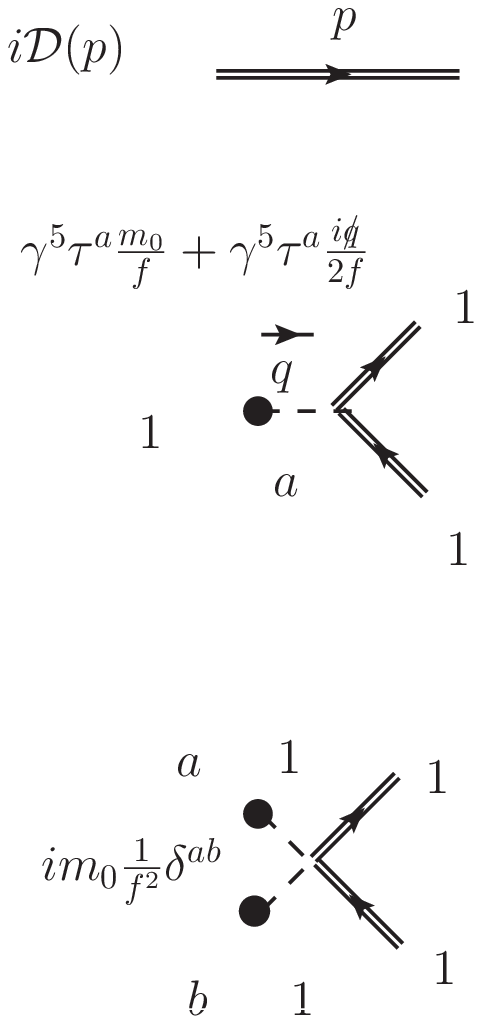}
\hspace{2cm}
\includegraphics[scale=0.40]{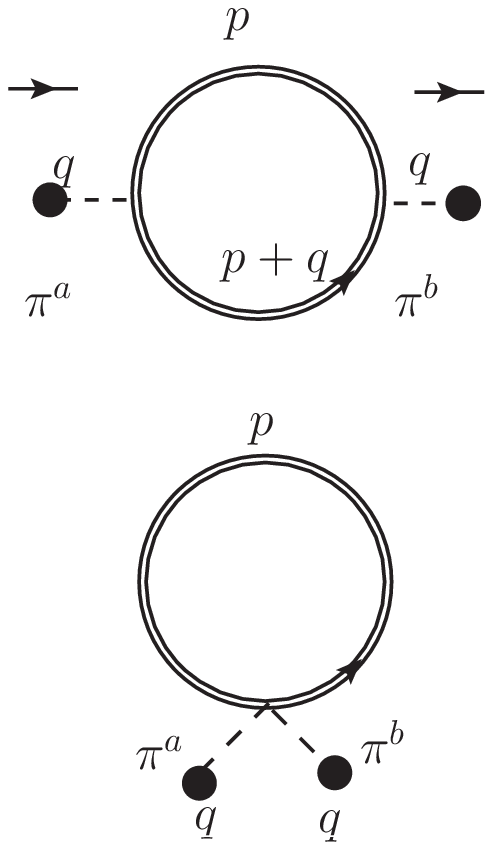}
\end{center}
\caption{The Feynman rules for the Lagrangian in \eqn{coupled} are shown on
 the left, where double lines denote quark propagators with Dyson-Schwinder
 resummation, and the blobs with attached short dashed lines stand for the
 non-propagating pion insertions. The one-loop Feynman diagrams for pion-pion
 correlators are shown in the right hand panel.}
\eef{feynman}

The Feynman rules for the Lagrangian in \eqn{coupled} are shown in
\fgn{feynman}, as are the Feynman diagrams for the pion two-point function.
Using these rules in the diagrams we find that the causal propagator is
\beq
 \int d^4x {\rm e}^{iq\cdot x}\left\langle\pi^a(0)\pi^b(x)\right\rangle
   = i\delta^{ab}\left[A(q^0)^2 - B (q^i)^2 - C + i\epsilon\right]^{-1},
\eeq{pioncor}
where $q$ and $x$ are Minkowski 4-vectors,
\beqa
\nonumber
 A &=& -\frac{i\N}{16f^2}\int\frac{d^4p}{(2\pi)^4}\,
    \tr[i{\cal D}_{11}(p)(i\gamma^5\gamma^0)i{\cal D}_{11}(p+q)
     (i\gamma^5\gamma^0)]\\
\nonumber
 B &=& \frac{i\N(d^4)^2}{16f^2}\int\frac{d^4p}{(2\pi)^4}\,
    \tr[i{\cal D}_{11}(p)(i\gamma^5\gamma^i)i{\cal D}_{11}(p+q)
     (i\gamma^5\gamma^i)]\\
 C &=& -\frac{i\N}{4f^2}\int\frac{d^4p}{(2\pi^4)}\left\{
    m_0^2 \tr[i{\cal D}_{11}(p)\gamma^5\,i{\cal D}_{11}(p+q)\gamma_5]
  + im_0 \tr[iD_{11}(p)]\right\},
\eeqa{integs}
and the trace is over Dirac components. The flavour trace gives the
factor of $\delta^{ab}$ in \eqn{pioncor}, and the colour trace is
subsumed into the factor of $\N$.

Since we are only interested in long-distance correlation functions of
pions, we need to take the limit $q\to0$. However, this can be done in
many ways. In the chiral limit, when $m_0=0$, all components of $q$
go to zero together, but one can take them to zero along lines of
varying $q^0/|\bf q|$. When $m_0>0$, which is physically relevant,
sending $q^0\to0$ while holding $|\bf q|$ fixed, gives us static
correlation functions. These are the same as in the Euclidean theory
\cite{Gupta:2017gbs}. Here we instead take the limit where $|\bf q|\to0$
keeping $q^0$ fixed, and only then take $q^0\to0$. In this limit we find
\beqa
\nonumber
  A &=& 
     \frac{\N m^2L}{32\pi^2(d^4)^3f^2}
    -\frac{\N m^2}{2(d^4)^3f^2}
           \int\frac{p^2dp}{(2\pi)^2 E^3}n_\F(E/T)\\
\nonumber
  B &=& 
     \frac{\N m^2L}{32\pi^2d^4f^2}
    +\frac{\N}{3d^4f^2}
           \int\frac{p^4dp}{(2\pi)^2 E^3}n_\F(E/T)\\
\nonumber
  C &=& 
     \frac{\N m^2m_0^2(1+L)}{16\pi^2(d^4)^3f^2}
    +\frac{2\N m_0^2}{(d^4)^3f^2}
           \int\frac{p^2dp}{(2\pi)^2 E}n_\F(E/T)\\
    &&
     -\frac{\N m^3m_0(1+L)}{16\pi^2(d^4)^3f^2}
    -\frac{2\N m_0m}{(d^4)^3f^2}
           \int\frac{p^2dp}{(2\pi)^2 E}n_\F(E/T)
\eeqa{ints}
where $L=\log((d^4)^2\mu^2/m^2)$, $\mu$ is the renormalization scale of
$\overline{MS}$, $n_F(E/T)$ is the Fermi distribution, and $E^2=m^2+p^2$.
In general causality imposes strong restrictions on the imaginary parts
of the propagator. In this case, however, since we work in the limit
$q\to0$, the imaginary parts vanish.  It is useful to note that the
logarithmic terms, which would survive in the limit $T\to0$, are the same
in Euclidean and real time. This non-trivial connection between real
time and Euclidean propagation of pions means that the recurrent idea
of inferring real time properties by measuring correlation functions
of composite operators on the lattice in the Euclidean time direction
using anisotropic lattices would be unproductive.

One can match the correlation function of \eqn{pioncor} with the integrals
in \eqn{ints} to an effective pion theory with a Minkowski Lagrangian
\beq
  L_f = -\frac12c^2 T_0^2\pi^2 + \frac12(\partial_4\pi)^2 
    - \frac12c^4(\nabla\pi)^2 - \frac{c^{41}}8\pi^4.
\eeq{pilag}
which includes all terms of dimension up to 4 (the coupling $c^{41}$
requires a separate computation which we do not present here). This is
valid for pion momenta $q\ll f$.  Note that a kinetic term for pions is
generated as usual after integrating out the quark fields.  The matching
involves a multiplicative renormalization of the pion field by $\sqrt
A$, so that the kinetic term is canonically normalized. Then one has the
identification $c^4=B/A$ and $c^2=C/A$. The usual definition of the pion
decay constant then gives $f_\pi=f/\sqrt A$.

\bef
\begin{center}
 \includegraphics[scale=0.25]{mpi.eps}
 \includegraphics[scale=0.25]{mkpi.eps}
 \includegraphics[scale=0.25]{fpi.eps}
 \includegraphics[scale=0.25]{upi.eps}
\end{center}
\caption{Comparison of various pion properties in a Euclidean theory
 against the corresponding property in real time, obtained using the
 Lagrangian in \eqn{qlag}. The couplings in the theory have been
 extracted from fitting to Euclidean lattice computations, and come
 with a band of uncertainty descended from statistical errors in the
 lattice measurement \cite{brandt}. The upper left panel contains the
 rest mass $m_r$ (pole mass in Euclidean), the upper right has the
 kinetic mass $m_k$ (which is compared to corresponding Euclidean
 quantity, \ie, the screening mass divided by $u_\pi$; this was divided
 by 4 for the purpose of display), the lower left
 shows $f_\pi$, and the lower right shows $u_\pi$.}
\eef{results}

The dispersion relations of a slowly moving particle with
dynamics obeying either \eqn{qlag} or \eqn{pilag} is
\beq
  E = \sqrt{m_r^2 + u^2p^2} = m_r + \frac{p^2}{2m_k} + \cdots
\eeq{disp}
where the rest mass $m_r=d^2T_0$ and $u=d^4$ for a quark, and
$m_r=\sqrt{c^2}T_0$ and $u_\pi=\sqrt{c^4}$ for a pion. The kinetic
energy term contains a different mass parameter, $m_k=m_r/u^2$. The
presence of the medium, and the consequent loss of the equivalence
of different frames, forces us to distinguish between the rest mass,
$m_r$, and the effective kinetic mass, $m_k$, even for particles which
are moving with relativistic speeds. The Euclidean version of the rest
mass $m_r$ has been called the pole mass.  The kinetic mass $m_k$ has no
exact analogue in the Euclidean, but may be compared with $m_s/u_\pi$,
where $m_s$ is the screening mass.  The values of both masses in real time
differ from those which are measured in Euclidean lattice computations.
A difference between Euclidean and real time bound state mass had been
noticed earlier in the Gross-Neveu model at large $N$ \cite{precursors}.
This computation shows that static screening phenomena in the effective
model of \eqn{qlag} are the same in the Euclidean and Minkowski theories,
whereas non-static correlators are different. As a result, the effective
model of \eqn{qlag} is a good tool for continuing the lattice computation
from Euclidean into real time.

Numerical results for the temperature dependence of the parameters of
the pion theory are shown in \fgn{results}. We note that the rest mass
in real time increases roughly linearly with $T$ below the cross over
temperature, $T_{co}$, and faster above that. At the lowest end of our
computation we find that $m_r=380\pm33$ MeV. This should be compared to
the lattice input, which tunes the quark masses such that $m_\pi=305\pm5$
at $T=0$. We note that this number was not an input to the fixing of the
couplings in \eqn{qlag}, since we do not expect these effective theories
to be valid at all temperatures. We also point out that the value of
$f_\pi$ decreases with increasing temperature. This is consistent with
the expectations of \cite{piondecay}. The behaviour of $m_k$ is closely
related to that of $u_\pi$. In the Euclidean computation $u_\pi$ falls
to zero in the chiral limit at the critical point, as a result of which
the pion correlations become unbounded. Even at the cross over at finite
quark mass, there is a decrease in the Euclidean $u_\pi$. On the other
hand, the real-time version shows no such decrease\footnote{We note that the
values of $u_\pi>1$ obtained in our computations do not violate causality,
since $dE/dp$ computed from \eqn{disp} remains less than unity for $p$ less
than the UV cutoff.}.

In summary, we have shown that pion propagation in Euclidean and Minkowski
space time at finite temperature involves two masses: the so-called pole
mass and screening mass in Euclidean, and a rest mass and kinetic mass in
real time. The relationship between the two pairs is non-trivial. This is
in strong contrast to the situation at $T=0$. Here we used an effective
theory of self-interacting quarks, defined through \eqn{qlag}, matched
to Euclidean lattice data to make the connection. The self consistency
of this procedure was tested by checking that the long-wavelength limit
of the static pion correlator obtained by taking $q^0\to0$ before taking
$|\bf q|\to0$ agrees with the Euclidean result. We also showed that in
real time $f_\pi$ decreases with increasing temperature, and that $u_\pi$
increases, in contrast to their trends on the lattice.

\end{document}